# Comments on "quantum oscillations in two-dimensional insulators induced by graphite gates"

Pengjie Wang, Sanfeng Wu[*]

Department of Physics, Princeton University, Princeton, New Jersey 08544, USA
[*]Email: sanfengw@princeton.edu

A recent Letter by Zhu et al [1] claimed a "unified mechanism" for quantum oscillations (QOs) in graphite-gated 2D devices based on studies of three cases (WTe$_2$, MoTe$_2$/WSe$_2$ heterobilayers and bilayer graphene (BLG)). A different microscopic mechanism for the graphite origin, was previously argued by the authors to explain qualitatively the same data (without the BLG case)[2]; the issues were pointed out by Wu[3]. The fact that different mechanisms were argued to explain the same data indicates that they lack solidity. Indeed, not only are the claims of [1] based on insufficient evidence in all three cases but also critical aspects are ignored, and hence their conclusions are unreliable.

(1) Zhu et al [1] argued that a rapid change of resistance $R$ under varying carrier density $n$, i.e., a large d$R$/d$n$, is necessary in the capacitive coupling mechanism to the observation of the graphite QOs in the 2D transport channel. However, they didn't mention that the strongest QOs observed in Wang et al[4] occurred at the most insulating plateau, where d$R$/d$n$ is mostly suppressed. Zhu et al[1] didn't observe QOs in that regime due to their limited device quality. Furthermore, correlations between WTe$_2$ QOs and graphite QOs appeared in the one branch observed in [1] does not provide a convincing way to tell the origin, as WTe$_2$ QOs can also induce changes in the graphite by the same argument. Wang et al [4] revealed much more information, including more QO branches, which are inconsistent with the "unified mechanism" proposed by [1].

(2) In the MoTe$_2$/WSe$_2$ moiré insulator, their key argument was that the QOs are absent in the TaSe$_2$-gated device[1]. This is however not a good contrast experiment since the TaSe$_2$-gated device exhibited a dramatically lower quality (e.g., significantly larger $R$ in the metallic regime, much weaker peaks and an opposite magnetoresistance compared to the graphite device). Ignoring this fact, which could cause the absence of QOs, Zhu et al claim that their limited data "*unambiguously confirms*" the graphite origin, an unreliable conclusion. New experiments are necessary to truly confirm or falsify it. Also ignored was the fact that the QO of the moiré insulator shows a distinct profile (zigzag) compared to that of WTe$_2$ [1], an indication of different origins.

(3) Zhu et al further claimed that they have observed the same graphite QOs in a gapped BLG device. Their arguments explicitly assumed that the bottom gate efficiently tunes the carrier density of the top gate, i.e., when the top gate voltage was changed by 0.09 V, the bottom gate voltage needs to be varied by 0.06 V (Fig. 4c in [1]) to maintain the same density in the top graphite. It implies a poor screening effect from the BLG, in striking contrast to their explicit assumption made in the WTe$_2$ case, where the carrier density in the top gate is strictly unchanged despite a huge variation of the bottom gate voltage (Fig. 2d in [1]). WTe$_2$ monolayer, even in the strongly insulating regime, was assumed to behave as a perfect screening layer like an excellent metal! This exotic assumption for WTe$_2$, completely opposite to that for the BLG, is again an issue ignored by

Zhu et al. Regardless of interpretations, the data presented in [1] has in fact shown that the WTe$_2$ monolayer insulator is distinct from the BLG insulator.

(4) Their arguments on BLG, if valid, call for re-examinations of previous studies of graphite-gated BLG devices, e.g., the earlier claim of fractional Chern states at high magnetic fields[5]. In the experiment, device capacitance was directly measured, where only those features with single gate dependence were attributed to the graphite gates[5,6]. However, (*i*) the single gate dependent features in capacitance measurements could come from single-gated physical area in the device, a possibility unrelated to the graphite QOs; (*ii*) In Zhu et al, the assigned graphite QO features depend on *both* top and bottom gates (Fig. 4c in [1]), implying that the discussion of graphite QOs in refs [5,6] was not sufficient. The inconsistency in discussing graphite features between Zhu et al[1] and refs [5,6] and between BLG and WTe$_2$ within Zhu et al[1] needs to be comprehensively addressed before reaching reliable conclusions in these reports.

It is important to examine the role of graphite in all these observations, which should be done carefully case by case. In the WTe$_2$ case, systematic discussions on various hypothetical explanations, including the graphite scenario, were included in the original paper[4]. New devices, e.g., high-quality ones without graphite gates, and new measurements are necessary to further test the hypotheses. Meanwhile, within a year since the publication of [4], the ingredients of the neutral fermion conjecture, highly exotic when first proposed, start to receive experimental and theoretical support. This includes the key presence of excitons[7–10]. A potential source of spin-charge separation, previously totally unexpected in WTe$_2$, has also recently been identified in a WTe$_2$ bilayer[11]. While the surprising observation in [4] is still calling for an convincing explanation, the study of the neutral excitations and Landau quantization of insulating WTe$_2$, initiated by[4,7] and echoed by the high field state of the bulk[12], are poised for advances.

## Acknowledgements

We acknowledge supports from NSF through a CAREER award DMR-1942942 and the Princeton University Materials Research Science and Engineering Center (DMR-2011750), ONR through a Young Investigator Award (N00014-21-1-2804) and the Eric and Wendy Schmidt Transformative Technology Fund at Princeton.